\documentclass[preprint2]{aastex}
\usepackage{color}
\usepackage{txfonts}
\usepackage{breqn}
\usepackage{psfrag,graphicx,natbib}
\shorttitle{First stars}

\shortauthors{Latif et al.}

\def\Rev. Mod. Phys{Rev.~Mod.~Phys}

\begin{document}

\title {The formation of massive Pop III stars in the presence of turbulence}

\author{M.~A.~Latif \altaffilmark{1}, D.~R.~G.~Schleicher \altaffilmark{1},  W.~Schmidt \altaffilmark{1}, J.~Niemeyer \altaffilmark{1}}
\affil{Institut f\"ur Astrophysik, Georg-August-Universit\"at, \\
 Friedrich-Hund-Platz 1, 37077 G\"ottingen, Germany}

\newcommand{\ch}[1]{\textcolor{red}{\textbf{#1}}}

\bibliographystyle{apj}

\begin{abstract}
Population III stars forming in the infant universe at z=30 heralded the end of the cosmic dark ages. They are presumed to be assembled in so-called minihaloes with virial temperatures of a few thousand K where collapse is triggered by molecular hydrogen cooling. A central question concerns their final masses, and whether fragmentation occurs during their formation. While studies employing Lagrangian codes suggest fragmentation via a self-gravitating disk, recent high resolution simulations indicated  that disk formation is suppressed. Here we report the first high-resolution large-eddy simulations performed with the Eulerian grid-based code Enzo following the evolution beyond the formation of the first peak, to investigate the accretion of the central massive clump and potential fragmentation. For a total of 3 halos, we see that a disk forms around the first clump. The central clump reaches $\sim10$~solar masses after 40 years, while subsequent accretion is expected at a rate of $10^{-2}$ solar masses 
per year. In one of these halos, additional clumps form as a result of fragmentation which proceeds at larger scales. We note that subgrid-scale turbulence yields relevant contributions to the stability of the protostellar disks. We conclude that the first protostar may reach masses up to $\rm 40-100~M_{\odot}$, which are only limited by the effect of radiative feedback.
\end{abstract}


\section{Introduction}

The first generation of stars, so-called Population III stars were formed at the end of the cosmic dark ages and their birth brought the first light into the cosmos. They emitted radiation and altered the dynamics of the Universe by heating and ionizing the gas \citep{2005SSRv..116..625C,2004ARA&A..42...79B}. They also polluted the Universe with metals and injected mechanical energy through stellar winds and supernova explosions \citep{2004ApJ...605..579Y,2005SSRv..116..625C,RevModPhys.85.809}. Consequently, they influenced the subsequent generation of structures through their feedback effects. To comprehend their formation is a matter of prime astrophysical interest.

According to the present paradigm of structure formation, Pop III stars were assembled in minihaloes with masses of few times $\rm 10^{5}-10^{6}~M_{\odot}$ \citep{2000ApJ...540...39A,2002ApJ...564...23B,2003ApJ...589..677O,2003ApJ...596L.135B,2004PASP..116..103B,2008ApJ...681..771M}. The collapse in these halos is induced by molecular hydrogen whose rotational and vibrational modes can be excited to lower temperatures of about a few hundred K. Previous studies showed that the first stars were massive with a mass range of about 30-300~M$_{\odot}$ and lived solitary lives \citep{2002Sci...295...93A,2007ApJ...654...66O,Yoshida08}. However, a new picture has emerged during the past few years. Recent simulations have shown that the protostellar disk fragments into multiple clumps which may lead to the formation of binary or multiple systems \citep{2009Sci...325..601T,2010MNRAS.403...45S,Clark11,Greif12}. 
Most of the simulations described above have been carried out using smoothed particle hydrodynamics (SPH), and the studies by \cite{2011ApJ...737...75G,Greif12} used the moving-mesh code AREPO \citep{2010MNRAS.401..791S}. All of these studies have thus been performed using Lagrangian methods. \cite{2009Sci...325..601T}, on the other hand, reported binary formation with a grid-based code, but as a result of two density peaks which happened to simultaneously collapse.

With grid-based methods, on the other hand, and the piecewise-parabolic method (PPM) available in Enzo, it is possible to study turbulent fragmentation much more accurately. In this context, previous numerical simulations demonstrated the requirement of a Jeans resolution of at least 32 cells to obtain converged turbulent energies \citep{2011ApJ...731...62F,2012ApJ...745..154T,2013MNRAS.tmp..551L}. The Jeans resolution employed in the previous fragmentation studies was however systematically lower. We note in particular that SPH simulations benefit from high resolution at the initial stage, but run out of resolution in the late stages of the collapse. \cite{2012ApJ...745..154T} reported the absence of a protostellar disk at high resolution per Jeans length, a result potentially inhibiting fragmentation via disk instabilities. Their study however explored only the formation of the first peak. It is thus essential to study fragmentation at high resolution per Jeans length, which goes  beyond the formation of 
the first peak. We further explore the impact of unresolved turbulence, which was found to have a strong impact on fragmentation in atomic cooling halos \citep{2013MNRAS.tmp..551L,2013arXiv1304.0962L}. Such turbulence may efficiently amplify magnetic fields during primordial star formation \citep{2010A&A...522A.115S,2010ApJ...721L.134S,2011ApJ...731...62F, 2012ApJ...745..154T,Schobera}.

We carry out large-eddy cosmological simulations by tracking the evolution beyond the formation of the first clump in minihaloes where cooling is regulated by molecular hydrogen. This allows us to study fragmentation at high resolution with the grid code Enzo after the formation of the first peak. We employ a fixed Jeans resolution of 64 cells per Jeans length throughout the evolution of the simulations and follow the collapse down to the scale of 0.1 AU for 40 years using the adaptive mesh technique. These are the first large eddy simulations to investigate the impact of SGS turbulence. This study will enable us to assess the impact of resolved and subgrid turbulence in the formation of first stars.

Our approach is the following. In section 2, we describe numerical methods and simulations setup. In section 3, we present our results. We summarize our main findings and discuss further implications in section 4.

\section{Simulations setup}

We present here simulations which are performed using a modified version of the adaptive mesh refinement code Enzo \citep{2004astro.ph..3044O} including a subgrid-scale (SGS) model which takes into account unresolved turbulence \citep{SchmNie06b}. The simulations are commenced with cosmological initial conditions generated from Gaussian random fields at $\rm z=99$ with a grid resolution of $\rm 128^{3}$ cells. The size of comoving periodic box is 300 kpc $\rm h^{-1}$ and the most massive halo lies at its center. We subsequently employ two additional nested refinement levels, each with a grid resolution of $\rm 128^{3}$ cells. In all, we use 5767168 particles to simulate the dark matter dynamics. This provides us a dark matter resolution of 70 $\rm M_{\odot}$. The parameters used for generating the initial conditions are taken from the WMAP seven years data \citep{2011ApJS..192...14J}. Additional 26 dynamical refinement levels are employed in the central comoving 26.7 kpc of the halo during the evolution of the simulation, giving a resolution of sub AU scales (in physical units). The refinement criteria used in these simulations are the same as in \cite{2013arXiv1304.0962L}. We mandated a resolution of 64 cells per Jeans length during the entire course of evolution. When the highest refinement level is reached, the thermal evolution becomes adiabatic. This approach enables us to follow the formation of structures beyond the formation of the first peak until a peak density of $\rm 1 \times 10^{-8}~ g~cm^{-3}$ is reached, which corresponds to 40 years.

Collapse in primordial minihaloes is triggered by $\rm H_{2}$ cooling which can bring the gas temperature down to a few hundred K. To model the primordial non-equilibrium chemistry, the rate equations of $\rm H$, $\rm H^{+}$, $\rm He$, $\rm He^{+}$,~$\rm He^{++}$, $\rm e^{-}$,~$\rm H^{-}$,~$\rm H_{2}$,~$\rm H_{2}^{+}$ are self-consistently solved in the cosmological simulations. We also include the effect of heating from the formation of molecular hydrogen via three-body reactions. This study was performed using the modified version of chemical solver developed by \cite{1997NewA....2..181A} and \cite{1997NewA....2..209A} (see \cite{2012ApJ...745..154T} for details).

To take into account the unresolved turbulence, large eddy simulations (LES) were performed employing the SGS turbulence model of \cite{SchmNie06b}. The latter is a mathematical model for turbulence based on a scale separation approach where resolved and unresolved scales are connected via an eddy-viscosity closure to transfer energy non-linearly across the grid scales. The turbulent viscosity is calculated on the grid scale and the subgrid scale turbulence is computed from self-consistently defined SGS energy generation and dissipation terms. On other hand, implicit large eddy simulations (ILES) use only the numerical dissipation coming from the discretization errors of the compressible fluid dynamics equations. The latter is the standard technique in computational astrophysics. Our previous studies describe further details of LES \citep{SchmNie06b,2009ApJ...707...40M,2011A&A...528A.106S}.

Our study consists of 3 distinct halos selected from different Gaussian random seeds. The results from the LES are compared with ILES. In total, we run 6 simulations.

\section{Main Results}

We have performed 3 LES and 3 equivalent ILES runs for three distinct haloes named A, B and C. They have masses of $\rm 1.3 \times 10^{5}~M_{\odot}$, $\rm 7 \times 10^{5}~M_{\odot}$ and $\rm 1 \times 10^{6}~M_{\odot}$ respectively. The collapse redshifts for these haloes are 21.7, 18.1 and 22.0. In the early phases of collapse gas falls in the dark matter potential, gets shock heated and leads to the nonlinear evolution. During the course of virialization, gravitational potential energy is continuously converted into kinetic energy of the gas and the dark matter. 

\begin{figure*}
 \vspace{-4.0cm}
\centering
 \hspace{1cm}
\includegraphics[scale=0.6]{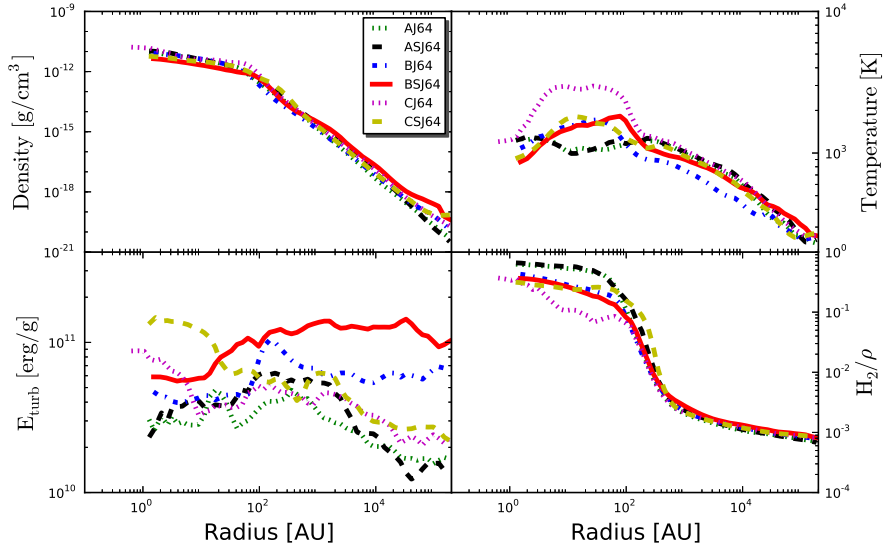}
\caption{The figure shows the radially binned spherically averaged radial profiles for 3 different halos. Each line style and color represents a simulation run as depicted in the legend (i.e, AJ64 is ILES run and ASJ64 is LES run, same for other haloes. This nomenclature is used in the rest of the figures). The upper panels of the figure show the density and temperature radial profiles. The turbulent energy (${1 \over 2} \rho v_{turb}^{2}$ for the definition of $\rm v_{turb}$ see equation 4 of \citep{2013MNRAS.tmp.1155L}) and H$_{2}$ fraction are depicted in the bottom panels of the figure.}
\label{fig0}
\end{figure*}

The properties of the halos at the maximum refinement level are shown in figure \ref{fig0}. The density profiles of different halos for LES and ILES runs agree very well and show $R^{-2.2}$ behavior at scales larger than 100 AU \citep{1998ApJ...508..141O}. The profile becomes flat in the center which is equivalent to the Jeans length. This behavior is consistent with earlier studies. Gas is initially heated up to its virial temperature and then subsequently cools by the rotational and vibrational modes of $\rm H_{2}$. At densities above $\rm 10^{8}~cm^{-3}$, $\rm H_{2}$ formation heating becomes effective and heats the gas to $\rm \geq$ 1000 K. The thermal evolution is the same for all halos at scales above 100 AU. Small temperature variations in the center may arise due to the differences in halo masses. The fraction of molecular hydrogen is low in the outskirts and close to one in the center which shows that the core of the halo becomes fully molecular. The amount of molecular hydrogen is slightly lower 
for massive halos and higher for the less massive halo. We attribute these differences to the self-regulation of molecular hydrogen due to the formation and collisional dissociation of $\rm H_{2}$ molecules. Furthermore,  the formation of $\rm H_{2}$ is very sensitive to the choice of three-body reactions \cite[for details see][]{2011ApJ...726...55T}. The specific turbulent energy is about $\rm 10^{11}~erg/g $ and the variations in turbulent energy are less than order of magnitude.

\begin{figure*}
\centering
\includegraphics[scale=0.3]{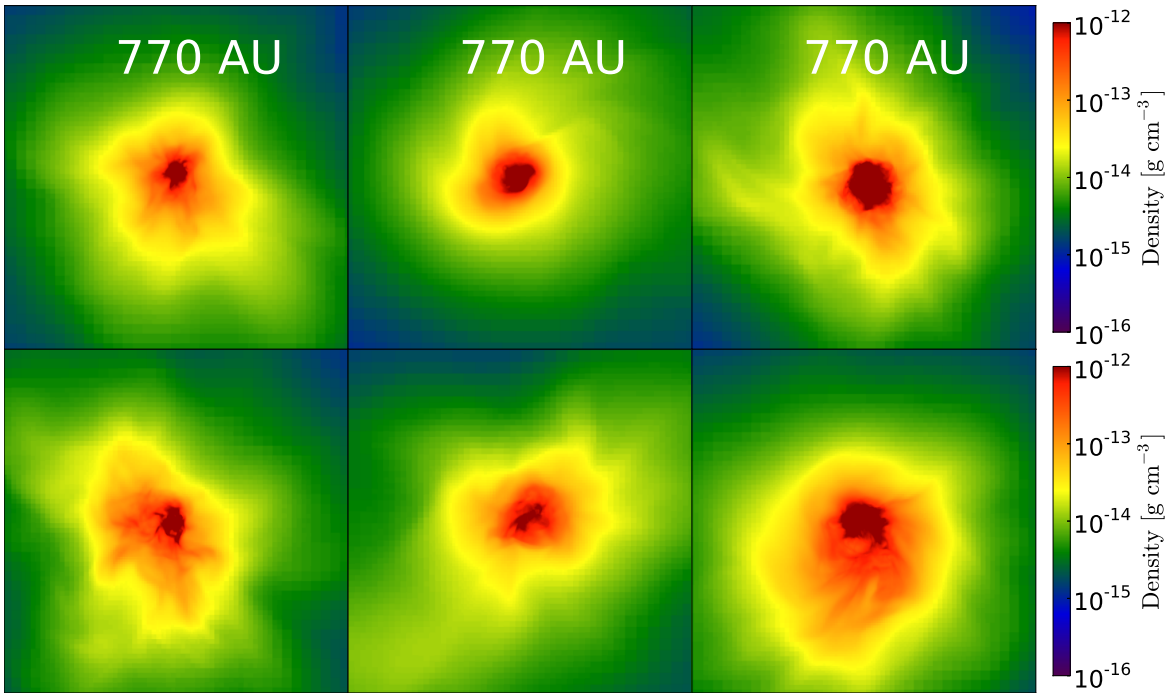}
\caption{The figure illustrates the state of simulations for 3 halos when the maximum refinement level is first reached. Density-weighted density projections are shown here for the central 770 AU of the halo for a fixed Jeans resolution of 64 cells. Top panels are ILES runs (from left to right halo A, B and C) while bottom panels are corresponding LES runs.}
\label{fig1}
\end{figure*}

The morphology of the haloes when simulations reach the maximum refinement level is shown in figure \ref{fig1}. At this stage of the collapse, the morphology is nearly-spherical for all haloes. There is no indication for the formation of a disk at this stage of collapse consistent with the results by \cite{2012ApJ...745..154T}. However, we see that in LES runs the morphology is different from ILES runs. Overall, the gas clouds are more extended in LES runs. This behavior is consistent with our previous study on more massive halos \citep{2013MNRAS.tmp..551L}.  

\begin{figure*}
\vspace{-2.0cm}
\hspace{-8.0cm}
\centering
\begin{tabular}{c}
\begin{minipage}{6cm}
\includegraphics[scale=0.3]{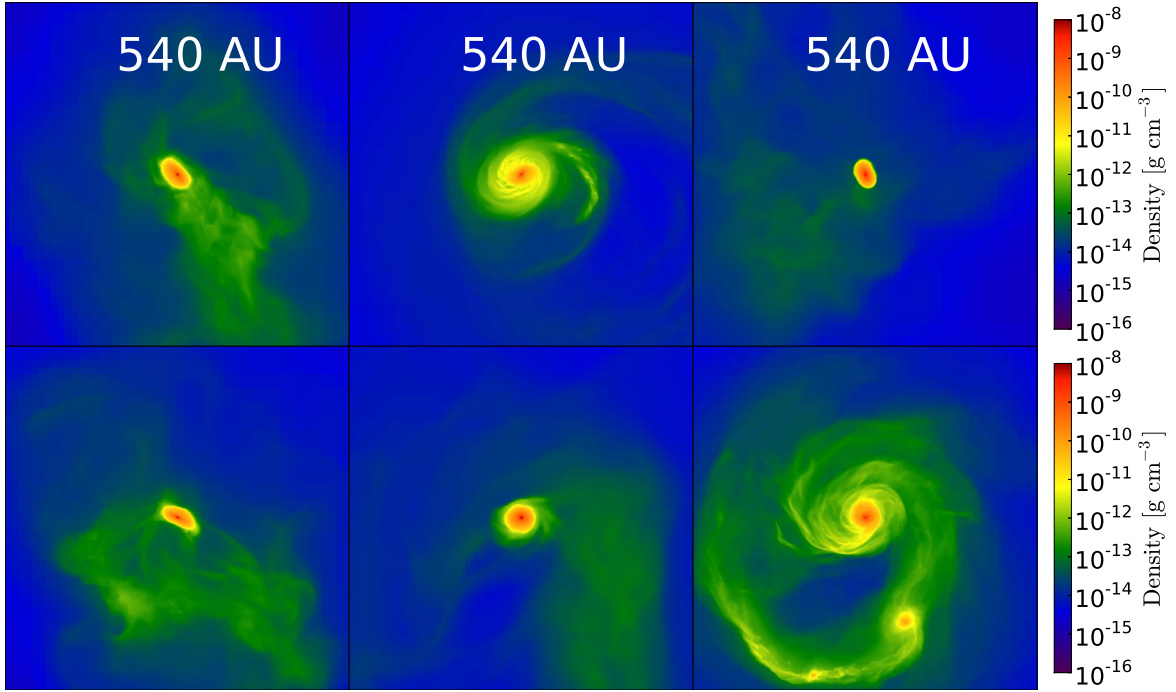}
\end{minipage}
\end{tabular}
\caption{Density projections for 3 halos (same as figure \ref{fig1}) for the central 540 AU at the final stage of the simulation which corresponds to 40 years.}
\label{fig2}
\end{figure*}

In order to follow the evolution beyond the first peak, we switch off the cooling above densities of $\rm 10^{-11}~g/cm^{3} $ to obtain an adiabatic evolution at higher densities when reaching the highest refinement levels. This approach allows us to study fragmentation without using sink particles. Turbulent accretion continues on the central clump and the final stage of the simulations is shown in figure \ref{fig2}. A disk is formed in the center of each halo for both LES and ILES runs. 
To further confirm the presence of a disk, we computed the eigenvalues of the inertia tensor and compared them with the state when the simulation reached the maximum refinement level. For halo A, the eigenvalues of the moments of inertia are almost equal (i.e., 1.74, 1.46, 1.39) when simulations reach the maximum refinement level. It indicates that the central clump is almost spherical at this stage and confirms our hypothesis that a disk is not formed yet. The eigenvalues of the moments of inertia after 40 years are 9.23, 6.23 and 6.49. The eigenvalues of two components of moments of inertia are equal while the value of the perpendicular component is about 1.6 times the values in disk plane. The expected values of the perpendicular component for a thin disk is twice the values in disk plane. We conclude that the disks formed here are thick in the vertical direction. Very similar results are found for the other haloes. In our simulated sample, one halo fragments into 3 clumps which may lead to the 
formation of multiple system. No fragmentation is observed in ILES runs.  There are also indications for the formation of further clumps in the rest of the halos outside the disk. Thus, we do not rule out the possibility of further fragmentation.

\begin{figure*}
\vspace{-2.0 cm}
\hspace{-9.0cm}
\centering
\hspace{9cm}
\includegraphics[scale=0.6]{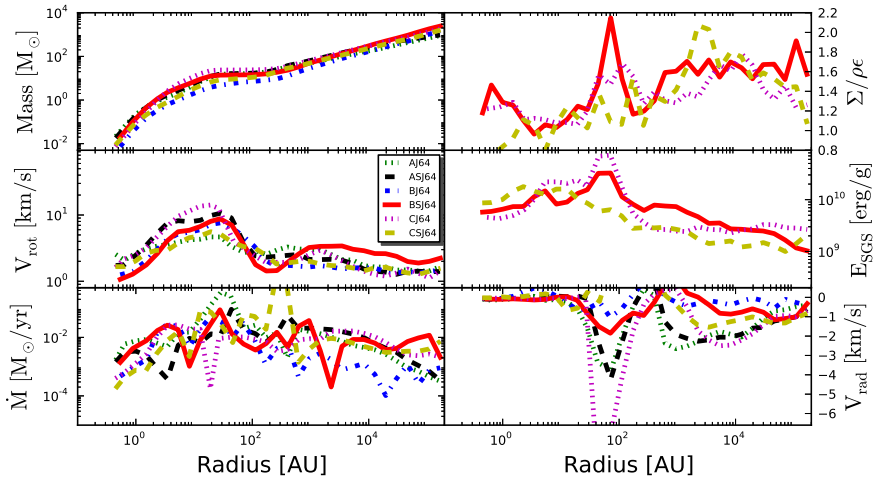}
\caption{Radial profiles for the physical properties of the disks centered on the peak density at the collapse redshifts. The mass, rotational velocity and mass accretion rate ($4\pi R^{2}\rho v_{rad}$) are shown in the left. The radial velocity, SGS energy and ratio of SGS turbulence production to dissipation is shown in the right.}
\label{fig3}
\end{figure*}

Radial profiles for the physical properties of the central disks are shown in figure \ref{fig3}. The typical rotational velocity is a few $\rm km/s$ for all disks.  It is low in the center, peaks around 10 AU and then there is a decline at larger radii following the Keplerian rotation. The radial velocity is a few $\rm km/s$ which indicates the infall of gas to the central disk. The average gas accretion rates on the central disks are a few times $\rm 10^{-3} ~M_{\odot}/yr$ which are typical for minihaloes. The mass radial profile increases sharply in the center, becomes flat and then linearly increases with radius. This behavior is observed for all runs. The masses of the central disks are about 10 M$_{\odot}$. The ratio of SGS turbulence production to dissipation shows they are in equilibrium in the center of the disk but production dominates in the surroundings excited by the gravitational instabilities \citep{2010A&A...520A..17K}. The subgrid scale energy is about comparable to thermal energy. It increases towards the center due to local production of turbulence during the gravitational collapse consistent with earlier studies \citep{2013arXiv1304.0962L}.

To quantify the role of turbulence, we followed the dynamics of gas compression employing a differential equation for the divergence of the velocity field $d=\nabla \cdot v$ \citep{2013arXiv1302.4292S}:
\begin{equation}
- {D d \over Dt}= 4 \pi G \rho_{0}\delta -\Lambda
\end{equation}

\begin{figure*}
\hspace{-2.0cm}
\centering
\begin{tabular}{c c}
\begin{minipage}{6cm}
\includegraphics[scale=0.4]{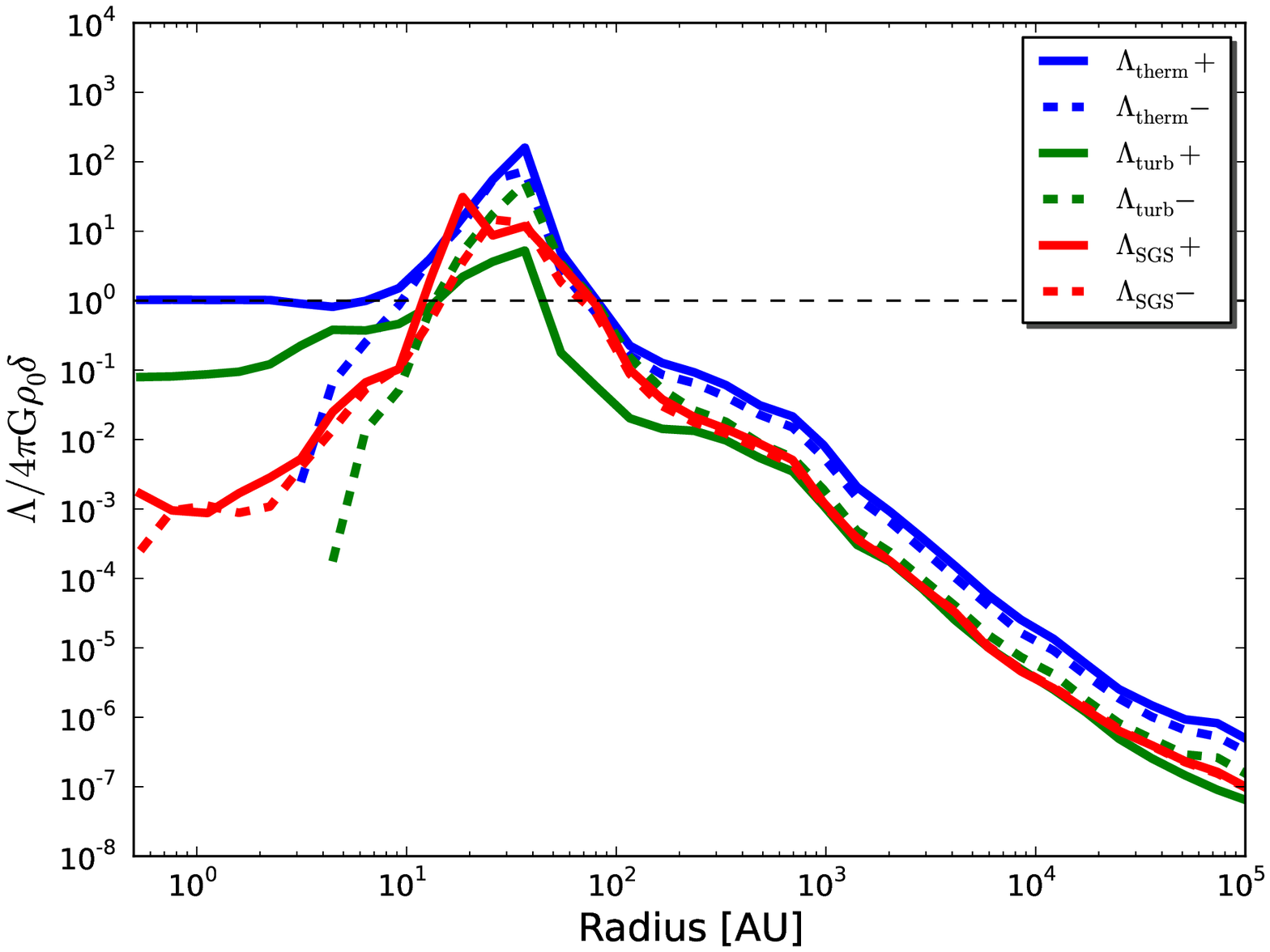}
\end{minipage} &
\hspace{1cm}
\begin{minipage}{6cm}
\includegraphics[scale=0.4]{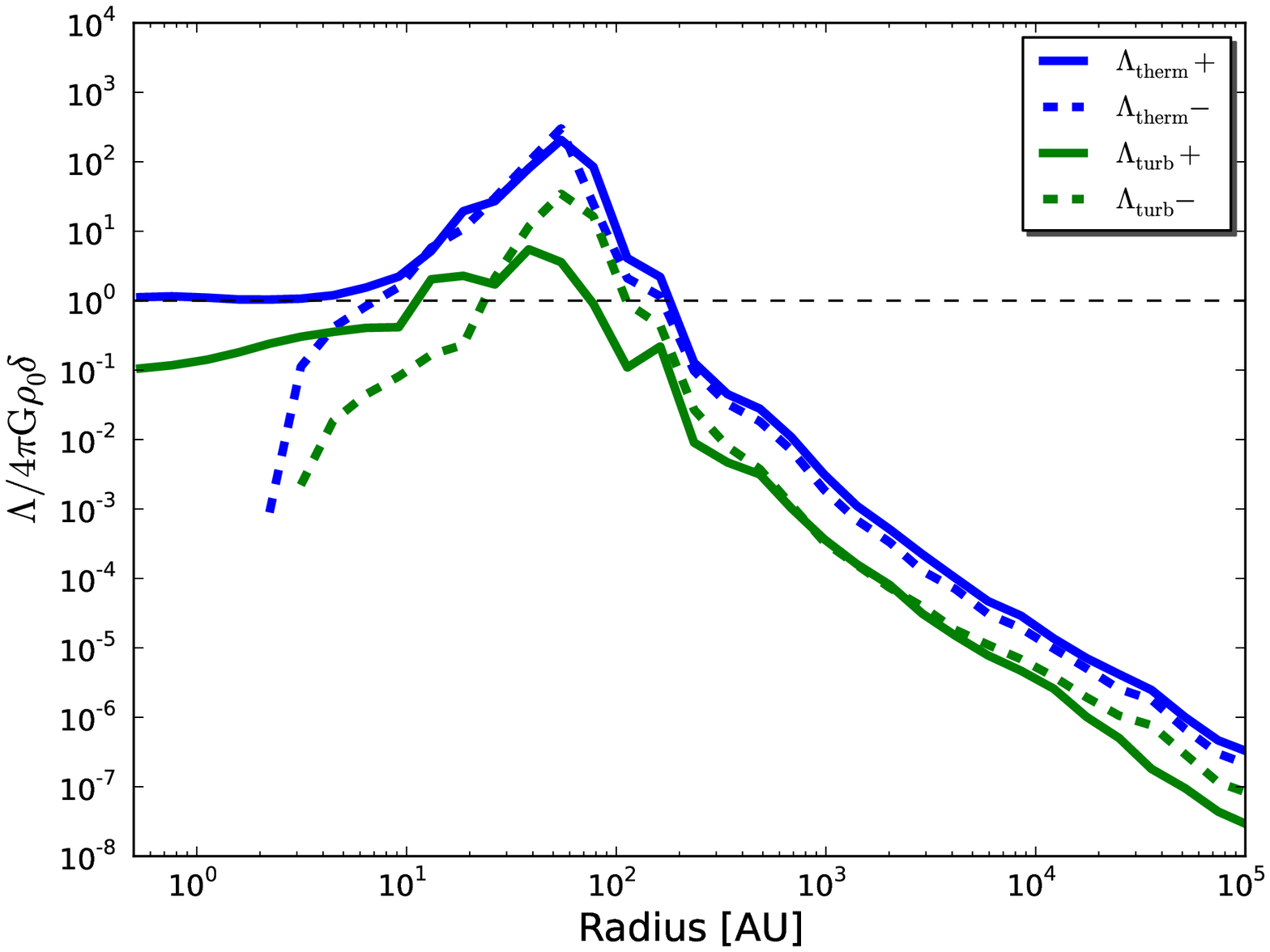}
\end{minipage} \\
\begin{minipage}{6cm}
\includegraphics[scale=0.4]{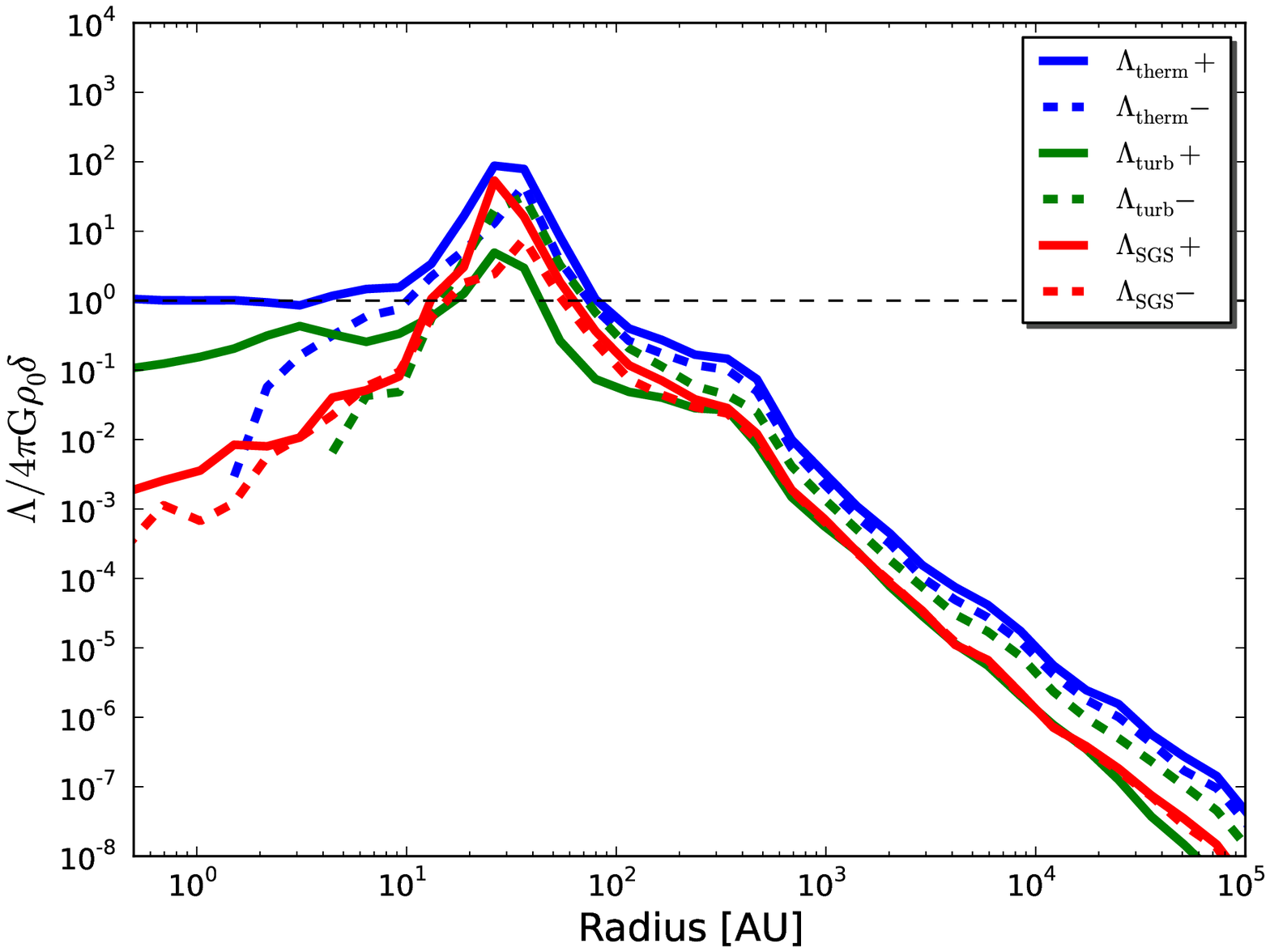}
\end{minipage} &
\hspace{1cm}
\begin{minipage}{6cm}
\includegraphics[scale=0.4]{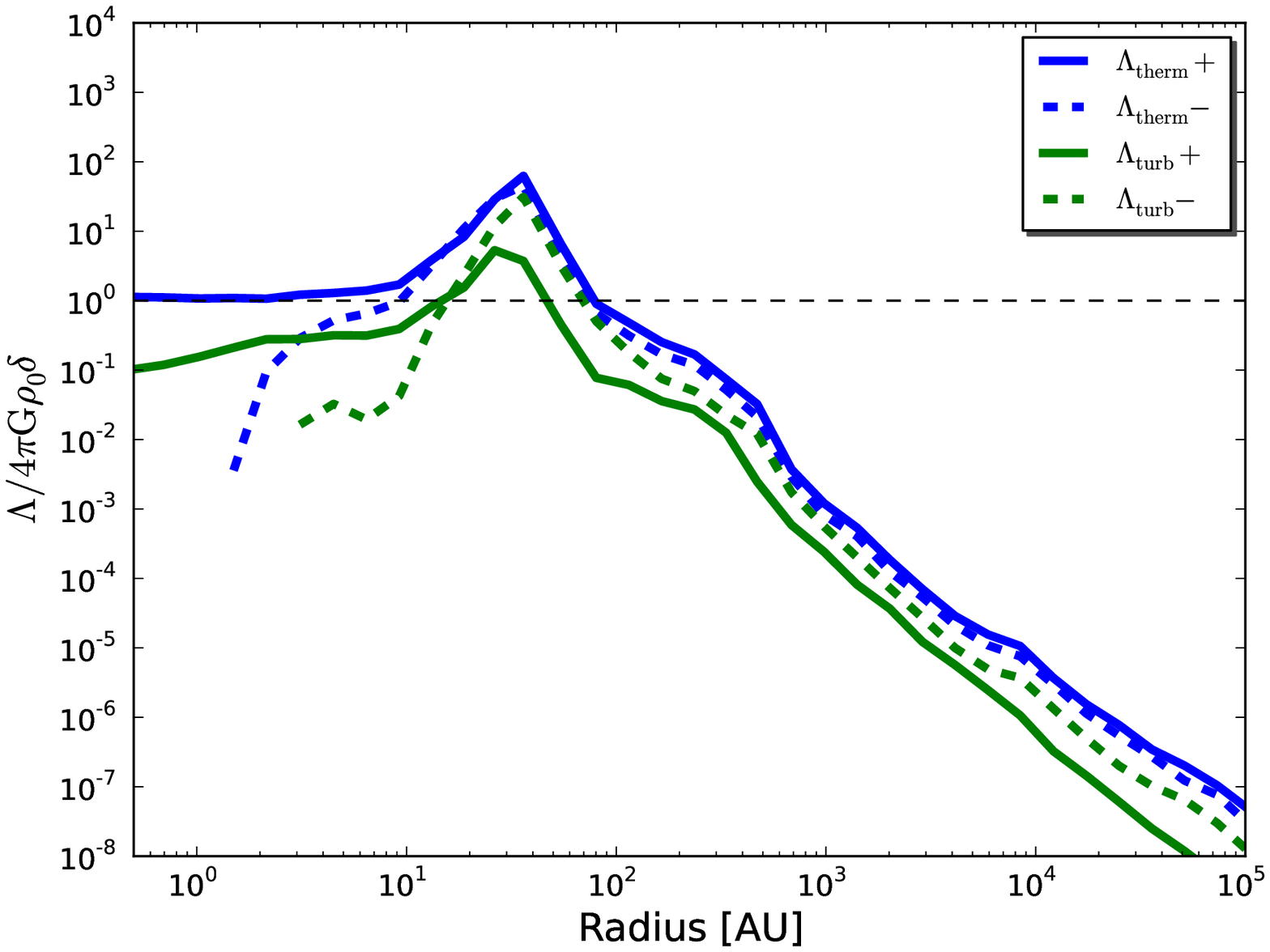}
\end{minipage} \\
\begin{minipage}{6cm}
\includegraphics[scale=0.4]{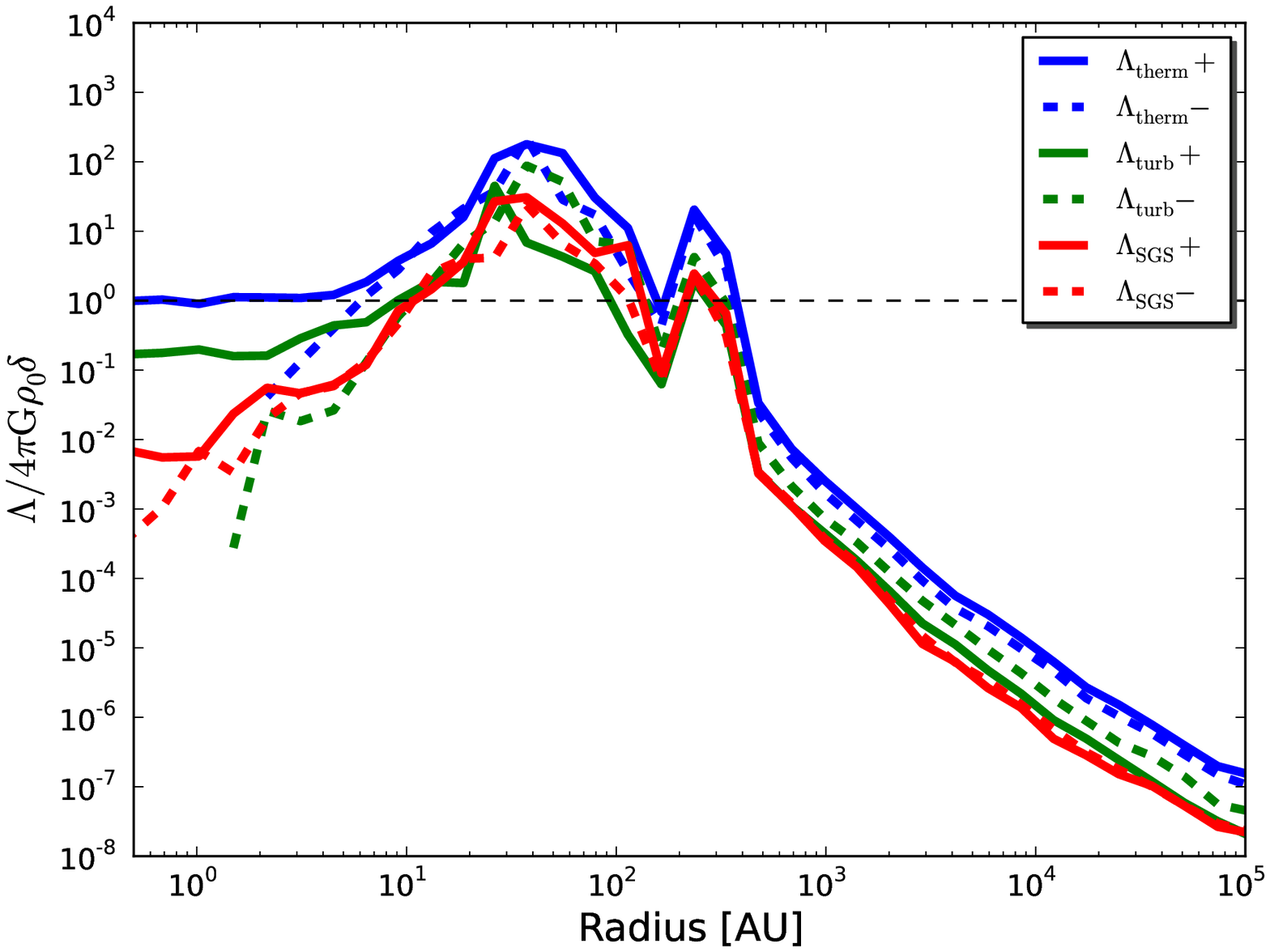}
\end{minipage} &
\hspace{1cm}
\begin{minipage}{6cm}
\includegraphics[scale=0.4]{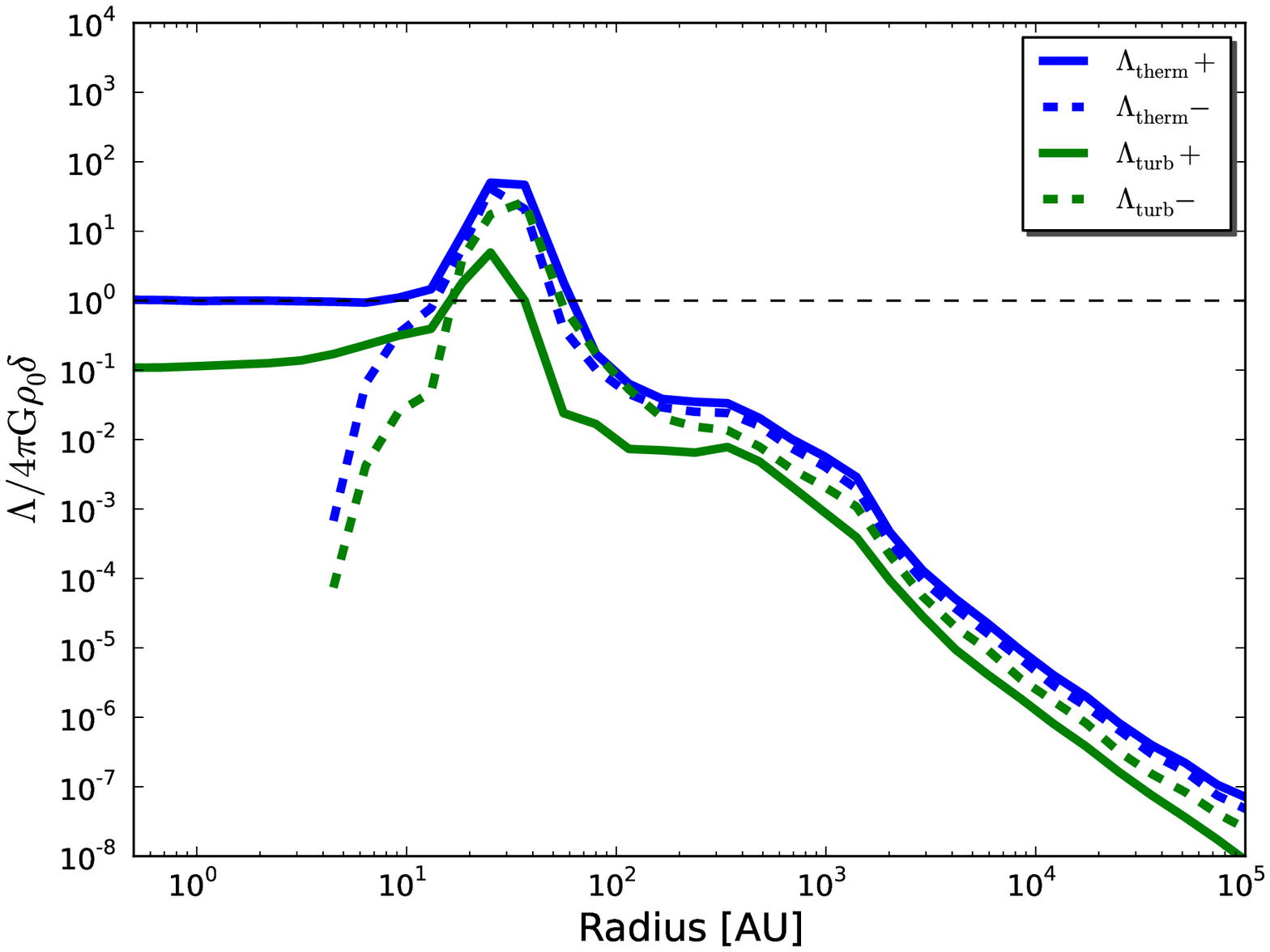}
\end{minipage}
\end{tabular}
\caption{Comparison of local support terms like thermal pressure, resolved turbulence and SGS turbulence for 3 haloes is shown here for LES (left)  and ILES (right) (top to bottom halo A, B and C). They are scaled by the gravitational compression. The positive components mean support against gravity while negative components aid the compression. Left side of the figure shows LES run and right side shows ILES runs.}
\label{fig4}
\end{figure*}

Here, ${D \over Dt}= {\partial \over \partial t} + v \cdot \nabla$, $\delta$ is the overdensity relative to the mean density $\rho_0$ and 
$\Lambda$ represents the local support terms against gravitational compression. $\Lambda$ receives contributions from thermal pressure, resolved turbulence and the SGS turbulent pressure. The contribution of local support terms like thermal pressure, resolved and subgrid turbulence scaled by gravitational compression (i.e., $4 \pi G \rho_{0}\delta$) are shown in figure \ref{fig4}. The compression by gravity dominates above 100 AU while  other supports terms remain sub-dominant. The impact of resolved and SGS turbulence becomes almost comparable to thermal pressure between 10-100 AU. The overall contribution of the positive component of SGS turbulence support becomes almost equal to the thermal pressure which acts against gravity and helps in the formation of disk.
The ratio of SGS-to-thermal support is about 30 \% which is a factor of 4 lower than found in atomic cooling halos \citep{2013arXiv1304.0962L}. We expect that positive support by SGS may become important at later stages of the collapse and particularly in the case of fragmentation on smaller scales. 
On the other hand, the negative component of resolved turbulence support aids the gravitational compression. In the very center of the halo, we have adiabatic core supported by thermal pressure. Similar behavior is observed for all haloes. 

\section{Discussion}

We present high resolution cosmological large-eddy simulations which track the evolution of high-density regions on scales of $0.1$~AU beyond the formation of the first peak in minihalos, and also explore the effect of subgrid-scale turbulence. This is the first study to explore fragmentation with a grid code employing the highest resolution per Jeans length in minihaloes which are predominantly cooled by molecular hydrogen. These simulations were performed for three distinct haloes. We compare the results of LES and ILES runs. In all, we carry out 6 simulations by following the collapse for 40 years beyond the first peak. Our results show that self-gravitating disks are formed  by turbulent accretion in all simulated haloes contrary to the previous study of \cite{2012ApJ...745..154T}. The masses of the disks are about $\rm 10~M_{\odot}$  at the final stage of our simulations and typical accretion rates are about $\rm 10^{-2}~M_{\odot}/yr$. Furthermore, fragmentation is observed in one out of 3 haloes 
which may lead to the formation of multiple systems. Moreover, the formation of additional clumps proceeds at larger scales. No fragmentation is found in ILES runs. The role of SGS turbulence becomes significant during the final stages of our simulation and may favor stable disks. Our findings suggest the formation of a relatively massive central protostar of $\rm  \geq 10~M_{\odot}$ via turbulent accretion. We further expect the turbulent accretion to continue until the radiative feedback from the protostar becomes effective.  

The Courant constraint limits the further evolution of simulations. We stopped our simulations after evolution of 40 years beyond the formation of the first peak when the simulations reached a peak density of $\rm 10^{-8}~cm^{-3}$. Previous studies suggest large variation in the fragmentation time scales. It occurs on relatively short time scales of about 10 years in \cite{Greif12} while in the simulations of \cite{Clark11} the disk becomes unstable after the evolution of 60-90 years. Fragmentation can thus still occur before the star reaches the main sequence. The time scale to reach main sequence is the Kelvin-Helmholtz time \citep[i. e., $\rm \sim 50,000$ years,][]{2011MNRAS.414.3633S}. We also note that it would be desirable to explore fragmentation further on sub-AU scales. Subsequent fragmentation may still occur, as the star keeps accreting at later times. 
In spite of this, it seems likely that at least one high mass star will form. The accretion luminosity is expected to become important during the early stages of a protostar formation. This was explored in detail by \cite{2011MNRAS.414.3633S,2012MNRAS.424..457S} finding that heating from accretion delays the fragmentation but does not completely prevent it. The protostar will continue to grow until the impact by UV feedback becomes significant, latter was examined by \cite{2011Sci...334.1250H} and \cite{2012MNRAS.422..290S}. They found that once the mass of the star exceeds  $\rm 8~M_{\odot}$ the Kelvin-Helmholtz (KH) time scale becomes shorter than the accretion time scale and the protostar passes through the so-called KH contraction phase. Consequently, the stellar luminosity increases as well as the UV flux. When photo-heating and photo-ionization by UV radiation becomes significant, the HII region breaks out from the disk above $\rm 20~M_{\odot}$, and further accretion will be suppressed. 

Here, we presumed that the halo is metal free but the presence of trace amounts of dust can change the thermal evolution by boosting the formation of molecules and may induce further fragmentation \citep{2009A&A...496..365C,2010MNRAS.402..429S,2012A&A...540A.101L,2013ApJ...766..103D}. In such cases, where the disks are more unstable due to the enhanced cooling, the additional SGS pressure will be even more relevant.


\section*{Acknowledgments}
The simulations described in this work used the Enzo code, developed by the Laboratory for Computational Astrophysics at the University of California in San Diego (http://lca.ucsd.edu). We acknowledge research funding by Deutsche Forschungsgemeinschaft (DFG) under grant SFB $\rm 963/1$, project A12 and computing time from HLRN under project nip00029. The simulation results are analyzed using the visualization toolkit for astrophysical data YT \citep{2011ApJS..192....9T}.


\end{document}